\documentclass[11pt]{article}
\usepackage[utf8]{inputenc}
\usepackage[margin=1.25in]{geometry}
\usepackage{amsmath, amsfonts, amssymb}
\usepackage[dvipsnames]{xcolor}
\definecolor{niceblue}{HTML}{236899}
\usepackage{hyperref}
\hypersetup{
    colorlinks=true,
    linkcolor=black,
    filecolor=black,
    urlcolor=niceblue,
    citecolor=niceblue,
    linkbordercolor = white
}
\usepackage{longtable}
\usepackage{array}
\usepackage{amsbsy}
\usepackage{epsfig}
\usepackage{bm}
\usepackage{xspace}
\usepackage{color}
\usepackage{lineno}
\usepackage{ragged2e}
\usepackage{comment}
\usepackage{multirow}

\usepackage{textcomp}
\usepackage[sb]{libertine}
\usepackage[varqu,varl]{inconsolata}
\usepackage[libertine,bigdelims,vvarbb]{newtxmath} 
\usepackage[cal=boondoxo]{mathalfa} 
\useosf 
\usepackage{setspace}

\definecolor{darkgrey}{HTML}{A9A9A9}

\modulolinenumbers[2]

\usepackage{booktabs}
\usepackage[round]{natbib}
\bibpunct{(}{)}{,}{a}{}{,}
\usepackage{authblk}
\usepackage{pdflscape}
\usepackage{setspace}

\date{}

\title{Introducing the generalized gamma distribution: a flexible distribution for index standardization}

\author[1*]{Jillian C. Dunic}
\author[2]{Jason Conner}
\author[1]{Sean C. Anderson}
\author[2]{James T. Thorson}

\affil[1]{Pacific Biological Station, Fisheries and Oceans Canada, Nanaimo, BC, Canada}
\affil[2]{National Oceanic and Atmospheric Administration, Alaska Fisheries Science Center, 7600 Sand Point Way NE, Seattle, WA, 98275}
\affil[*]{corresponding author: jillian.dunic@dfo-mpo.gc.ca}

\begin{document}
\maketitle

\begin{abstract}
Fisheries scientists use regression models to estimate population quantities, such as biomass or abundance, for use in climate, habitat, stock, and ecosystem assessments.
However, these models are sensitive to the chosen probability distribution used to characterize observation error. 
Here, we introduce the generalized gamma distribution (GGD), which has not been widely used in fisheries science.
The GGD has useful properties: (1) it reduces to the lognormal distribution when the shape parameter approaches zero; (2) it reduces to the gamma distribution when the shape and scale parameters are equal; and (3) the coefficient of variation is independent of the mean.
We assess the relative performance and robustness of the GGD to estimate biomass density across different observation error types in a simulation experiment.
When fit to data generated from the GGD, lognormal, gamma, and Tweedie families, the GGD had low bias and high predictive accuracy.
Finally, we fit spatiotemporal index standardization models using the R package sdmTMB to 15 species from three trawl surveys from the Gulf of Alaska and coast of British Columbia, Canada.
When the Akaike information criterion (AIC) weight was compared among fits using the lognormal, gamma, and Tweedie families the GGD was the most commonly selected model. 

\end{abstract}
    

\section{Introduction}

A primary challenge of fisheries research is the accurate and precise estimation of fish population characteristics (e.g., population size and variance) for inclusion in assessments of fish stocks, habitat, and the status of marine ecosystems. 
Regression models provide researchers with a versatile tool for fitting predictions to various data sources (e.g., observations from fisheries-independent surveys), but these models depend on the skill of the researcher to specify the ``best'' probability distribution for the data. 
We define ``best'' as a distribution with the fewest parameters (to avoid overfitting) that conforms to current understanding of a fish population and that fits observed data well over its range (i.e., residuals are small and are homoscedastic). 
The selection of theoretical distributions may lead to sizable differences in the estimation of the scale of a fish population \citep{thorson2021scale}. 
While fisheries researchers commonly apply the lognormal or gamma distributions \citep{maunder2004} to model variables observed with a continuous positive response, distributions with more than 2 parameters may be overlooked or ignored.

\subsection{Introducing the generalized gamma distribution}

The generalized gamma distribution (GGD) is a 3-parameter distribution that provides multiple flexibilities for modelling fisheries data.
\citet{amoroso1925} described a continuous function with four parameters to analyze the distribution of income data. 
The Amoroso distribution has four parameters (a location parameter, scale parameter, and two shape parameters) and includes special cases for several useful distributions, such as the lognormal and gamma distributions \citet{crooks2019FieldGuide}, which are commonly used by fisheries researchers.
\citet{stacy1962} removed the location parameter from the Amoroso distribution and described a generalization of the gamma distribution with three parameters, which also includes the special case for the lognormal distribution.
While the properties of the Amoroso and Stacy parameterizations are useful for biological analyses, the parameterization in \citet{stacy1962} does not lend itself to maximum likelihood estimation (MLE) methods \citep{prentice1974}.
\citet{prentice1974} applied two transformations to the \citet{stacy1962} parameterization: a square root transformation of shape parameter $k$ and a log transformation of the probability density function (PDF).
This parameterization improves numerical stability when estimating $Q \to 0$ and allows for $Q < 0$ \citep{jackson2016flexsurv}.
As of the writing of this paper, the generalized gamma distribution has previously been implemented in R packages flexsurv \citep{jackson2016flexsurv}, ggamma \citep{saldanha2019ggamma}, and \emph{gamlss} \citep{rigby2005gamlss}) and in the Python library SciPy \citep{2020SciPy}, making the VAST \citep{thorson2019vast} and sdmTMB \citep{anderson2024sdmTMB} R package implementations described in this paper the first to include this distribution for fisheries analysis.

\subsection{Fisheries applications}\label{fisheries-applications}

Because the GGD has three parameters, it provides more flexibility in fitting fisheries data than the more commonly used lognormal or gamma distributions, which is advantageous if the decrease in estimation error offsets the increased degrees of freedom (and resulting increase in expected loss for out-of-sample relative to in-sample prediction) compared to distribution families with fewer parameters.
Fisheries researchers may select the GGD for model-based estimates (including spatial, temporal, or spatiotemporal models) of the mean and variance of biomass or abundance of fish species, to standardize indices of abundance, or to specify error distributions in fish stock assessments.
Another benefit of the GGD is that MLE methods are able to estimate distribution parameters for lognormal and gamma distributions, neither of which have special cases for the other.

Here, we compare the performance of models that use the lognormal, gamma, Tweedie, and GGD in (1) a simulation experiment and (2) when applied to time series of observations of effort-standardized catch (i.e., catch per unit effort, hereafter CPUE) from fishery-independent surveys in the Gulf of Alaska and off the coast of British Columbia, Canada.
We also select three species that exhibit different life history and population structure attributes and examine how models fit to these data differ.
By demonstrating the efficiency of the GGD and its ability to model both the lognormal and gamma distributions, we propose that fisheries researchers explore using the GGD for modelling fish populations.

\section{Methods}

\subsection{GGD parameterization}

\begin{figure}[htb]
    \centering
    \includegraphics[width=1\linewidth]{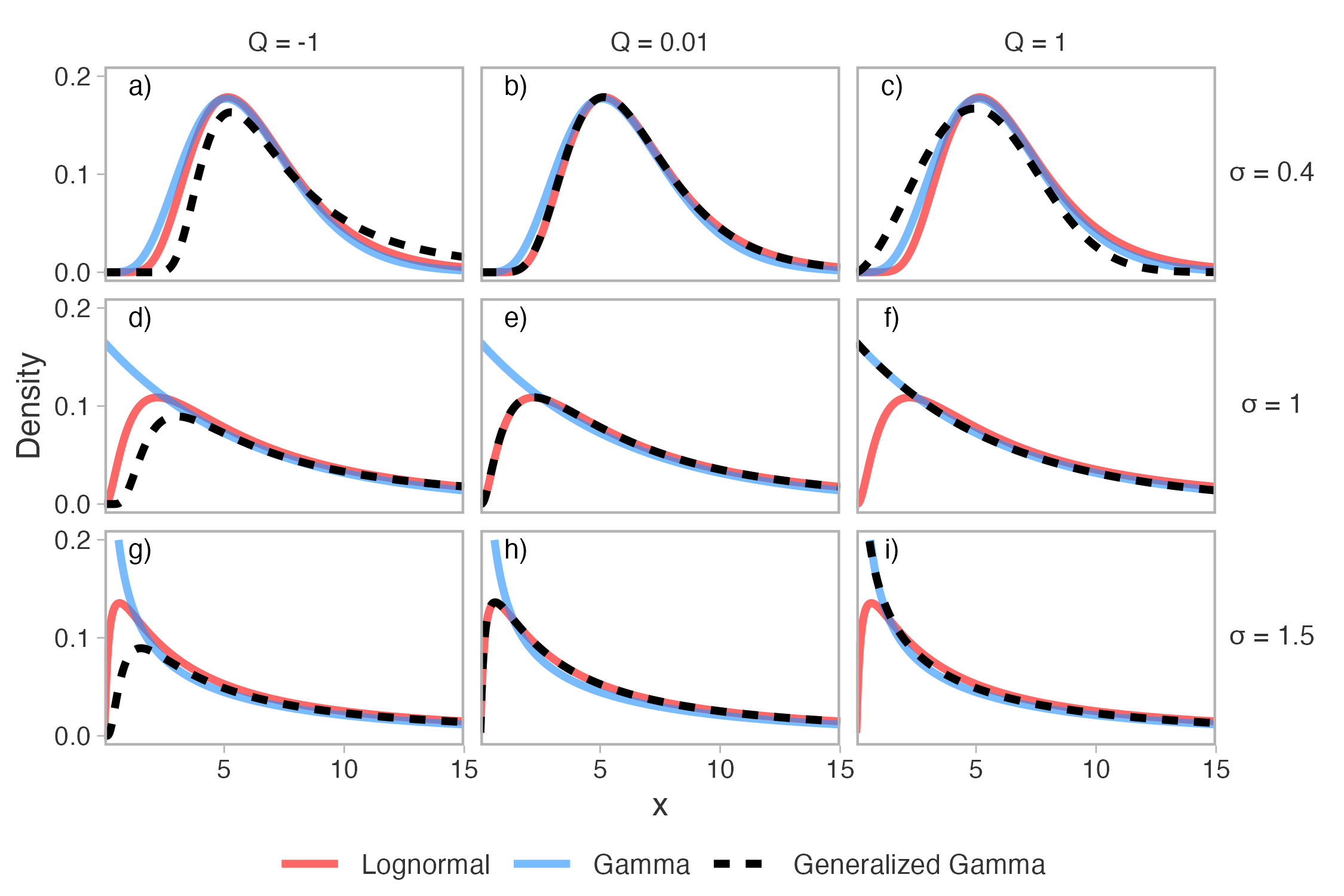}
    \caption{Comparison of densities for three distributions across three values of the generalized gamma Q shape parameter (columns) and three scale parameter values $\sigma$ (rows). The lognormal (red) and gamma (blue) are the same across rows. The location parameter $\mu$ is fixed at 1.8.}
    \label{fig:compare-dists}
\end{figure}

We used a mean-CV form of the \citet{prentice1974} GGD parameterization, where we use the mean ($\bar{x}$) instead of location ($\mu$) (Table \ref{tab:table1}). 
Internally, the generalized linear model linear predictor $\eta$ (i.e., $\log(\bar{x})$) is converted into $\log(\mu)$ as
\begin{equation}
\log(\mu) = \eta - \log \Gamma \left(\frac{k \beta+1}{\beta} \right) + \log \Gamma(k) + \frac{\log(k)}{\beta},
\end{equation}
\noindent
where $k = Q^{-2}$, $\beta = Q / \sigma$, and $\log \Gamma$ is the log gamma function.
Special cases of the GGD are that it reduces to the lognormal distribution when \(Q \to 0\) (Figure~\ref{fig:compare-dists}b,e,h) and it reduces to the gamma distribution when $\sigma = Q$ (e.g., Figure~\ref{fig:compare-dists}f).
Thus, with the addition of one parameter, relative to the lognormal or gamma distributions, the GGD can model either distribution, a distribution in between the two, a distribution with lighter tails than the gamma, or a distribution with heavier tails than the lognormal. 
An additional special case for the GGD is the Weibull distribution when \(Q = 1\) (Figure~\ref{fig:compare-dists}c,f,i), and (Figure~\ref{fig:compare-dists}f).
Finally, the coefficient of variation for the GGD is a function of the $\sigma$ and $Q$ parameters, unrelated to $\bar{x}$ (Table \ref{tab:table1}).

\begin{table}[htb]
    \centering
    \caption{
Selected properties of the generalized gamma distribution. 
We define $\mu$ as the location parameter, $\sigma$ as the scale parameter, and $Q$ as the family or shape parameter. 
In this form of the probability density function (pdf), $\mu > 0$ and $\sigma > 0$, and $Q \in (-\infty, +\infty)$.
We further define two temporary variables to simplify notation:
$\beta = Q / \sigma$ and $k = Q^{-2}$. The symbol $\Gamma$ is the gamma function.
}
    \begin{tabular}{ll}
    \toprule
    Property & Equation \\
    \midrule
        mean & 
        $\Gamma \left(\frac{k\beta + 1}{\beta}\right) 
        \cdot \frac{1}{\Gamma k}
        \cdot e^{\mu - \frac{\log(k)}{\beta}}$ \\
        
         variance & 
         $\left(e^{\mu-\frac{\log(k)}{\beta}}\right)^2 
         \cdot \left[ \frac{\Gamma\left(\frac{k\beta+2}{\beta}\right)}{\Gamma(k)} -
         \left(\frac{\Gamma\left(\frac{k\beta+1}{\beta}\right)}{\Gamma(k)} \right)^2 \right]$ \\

         CV &
         $\left(\Gamma \left(\frac{k\beta + 1}{\beta}\right)\right)^{-1}
         \cdot 
         \sqrt{\Gamma \left( \frac{k \beta + 2}{\Gamma k} \right) \cdot \Gamma k}$ \\
         
         pdf collapses to gamma when     & $Q = \sigma$ \\
         
         pdf collapses to lognormal when & $Q \to 0$ \\
        \bottomrule
    \end{tabular}
    \label{tab:table1}
\end{table}

\subsection{Simulation testing}

We assess the relative performance and robustness of the GGD to estimate biomass density across different observation error types. 
In a factorial-simulation experiment we examine the goodness-of-fit, bias in estimates, and predictive accuracy of the GGD, lognormal, gamma, and Tweedie families. 
The Tweedie family is not a special case of the GGD but we have included it in our analyses because it is widely used in fisheries applications \citep{foster2013} and produces index scale estimates that are similar to design-based indices \citep{thorson2021scale}.

Using sdmTMB, we simulated 10000 points across a 100 $\times$ 100 cell grid using a spatial model of the form: 
\begin{equation} \label{cross-sim-eqn}
\begin{aligned}
\mathbb{E}[y_{\bm{s}}] &= \mu_{\bm{s}},\\
\mu_{\bm{s}} &=
g^{-1} \left(\alpha + \omega_{\bm{s}} \right),\\
\bm{\omega} &\sim \operatorname{MVNormal} \left( \bm{0}, \bm{\Sigma}_\omega \right),
\end{aligned}
\end{equation}
\noindent
where the expected value $\mathbb{E}[y_{\bm{s}}]$, of catch $y_{\bm{s}}$ at spatial coordinates $\bm{s}$ is the mean at that point in space, given by $\mu_{\bm{s}}$. The mean is equal to an inverse link function $g^{-1}$ applied to the observed catch $\alpha$ plus a spatial Gaussian Markov random field (GMRF) value $\omega_{\bm{s}}$. 
We used a Gaussian Markov Random Field with a marginal standard deviation of 1 and range \citep[distance at which correlation decays to $\approx$ 0.13;][]{lindgrenExplicitLink2011} of 0.5.
From the simulated expected values, we sampled observations for 500 locations and added observations errors from one of our four families. 

Our four families used either a single linear predictor and the Tweedie distribution or delta-models \citep{aitchison1955, maunder2004} (also called ``hurdle'' models), which had two linear predictors (one predicting encounter probability and one predicting catch conditional on encounter).
We used a form of the Tweedie distribution \citep{tweedie1984} (also known as the compound Poisson Gamma), that can scale between the Poisson (as the power parameter $p \to 1$) and the gamma ($p \to 2$); in this simulation we used Tweedie $p = 1.5$.
For the delta-models, we simulated encounters (mean encounter probability = 0.5), using a Bernoulli distribution and a logit link. 
The positive catch rates (mean catch rate = 1) for the delta-models were sampled from one of gamma, lognormal, or the generalized gamma distribution.
We standardized the scale parameter of each observation error distribution so that the CV of the observation error was 0.95, which was similar to the mean CV observed across species in the multi-stock analysis.
We also included a sequence of values of Q values (Q = -2, -1, -0.5, -0.001, 0.001, 0.5, 0.95, 1, 2) to assess the families against data with a range of heavy to light tails, to assess the ability for the GGD models to estimate Q, and illustrate the equivalence of the GGD to the lognormal at $Q = 0$ and to the gamma when $Q = \sigma$ (here at $Q = 0.95$).

We then fit the model specified in equation \ref{cross-sim-eqn} to each simulated dataset using the delta-GGD, delta-lognormal, delta-gamma, and Tweedie families.
Using these model fits, we summed density predictions across a prediction grid to generate an area-weighted biomass index \citep[e.g.,][]{thorson2015geo}, using a generic bias-correction estimator \citep{thorson2016bias}. 
We repeated this sampling, model fitting, and index calculation 1000 times, each time generating a new GMRF, new sampling locations, and new observation error.
For each simulation iteration, we calculated the relative error (RE) of the predicted index value ($\hat{I}$) compared to the true index ($I_\mathrm{true}$), $\mathrm{RE} = \left(\hat{I} - I_\mathrm{true}\right) / I_\mathrm{true}$, and calculated the marginal AIC \citep[Akaike Information Criterion;][]{akaike1973} weight. 

To illustrate goodness of fit in the tails of the distributions, we performed an additional single simulation with an increased sample size.
We simulated observations at 2000 points, fit each of the four models, and then calculated randomized quantile residuals \citep{dunnRandomizedQuantile1996} for the full dataset (Tweedie) or the positive linear predictor (all other families) making sure to sample random effects from their approximate (multivariate normal) distribution rather than use their empirical Bayes estimates \citep{waagepetersen2006, thorson2024book}.

\subsection{Multi-stock analysis}\label{multi-stock-methods}

We compared the performance of the GGD relative to the lognormal, gamma, and Tweedie when applied to trawl survey data from the Gulf of Alaska and off the coast of British Columbia, Canada. 
We fit models to 15 species from three regions: the Gulf of Alaska (GOA), Hecate Strait and Queen Charlotte Sound (HS-QCS), and West Coast Vancouver Island (WCVI).
These species occurred in each of the three regions and were well-sampled in the HS-QCS and WCVI regions, meaning they were encountered in at least 20\% of samples over time.
We used this encounter cutoff for the HS-QCS and WCVI regions because they have much smaller spatial coverage and lower annual data availability than the GOA (Figure \ref{fig:sampling-diff-supp}). 
Time series spanned 18--20 years in BC survey regions (10--12 sampled years) and 30--33 years (15 sampled years) in the Gulf of Alaska. 
For the HS-QCS and WCVI surveys, we used a SPDE mesh with a ``cutoff'' (minimum triangle edge length) of 8 km \citep{fmesher}, and for the GOA surveys we used a cutoff value that resulted in a mesh with 500 knots ($\approx$ 26.5 km). 
We fit the following model for each family, with a constant spatial GMRF and independent GMRFs by year as: 
\begin{equation}
\begin{aligned}
\mu_{\bm{s},t} &=
g^{-1} \left(\alpha_t + O_{\bm{s},t} + \omega_{\bm{s}} + \epsilon_{\bm{s},t} \right),\\
\bm{\omega} &\sim \operatorname{MVNormal} \left( \bm{0}, \bm{\Sigma}_\omega \right),\\
\bm{\epsilon}_{t} &\sim \operatorname{MVNormal} (\bm{0}, \bm{\Sigma}_{\epsilon}),
\label{eq:stocks}
\end{aligned}
\end{equation}
\noindent
where the expected mean $\mu_{\bm{s},t}$ at spatial coordinates $\bm{s}$ at time (year) $t$, is equal to an inverse link function $g^{-1}$ applied to the linear combination of observed catch $\alpha_t$, offset $O_{\bm{s},t}$ for log area swept, a spatial GMRF value $\omega_{\bm{s}}$, and spatiotemporal GMRF value $\epsilon_{\bm{s},t}$. 

As in the simulation experiment, we fit the models as either a single linear predictor and the Tweedie distribution or a delta-model that had two linear predictors (one predicting encounter probability and one predicting catch conditional on encounter) for the lognormal, gamma, and GGD families. 
We considered models to have converged if the maximum absolute marginal log-likelihood gradient with respect to all fixed effect parameters was $< 0.005$, the Hessian was positive definite, no random field marginal standard deviations were on a boundary (defined as $< 0.01$), and the mean coefficient of variation in the estimated index was $< 1$.
If the model failed to converge, we checked if any of the spatial or spatiotemporal random fields had collapsed to zero (i.e., the marginal field SD $< 0.01$), and if they had, we omitted those random fields (set to zero) and refit the model.
In a few cases where the delta-GGD model did not converge (three stocks), we refit models in two ways: we tested the use of a weak prior that constrained the year estimates, where $\alpha_{t} \sim \operatorname{N}(0, 30^{2})$ or we scaled the catch variable by 0.01 kg because the estimation of the GGD may be sensitive to scale and/or the starting values used for $sigma$. 
We include these results in the Supporting Information to offer readers practical options for troubleshooting convergence issues when fitting the GGD model.
We compared the AIC weights ($w_i$) across the all of the fitted models $R$ within each species-region as:  

\begin{equation}
\begin{aligned}
w_i = \frac{\exp\left(-\frac{1}{2} \Delta_i\right)}{\sum_{r=1}^{R} \exp\left(-\frac{1}{2} \Delta_r\right)} \\
\Delta_i = \text{AIC}_i - \text{AIC}_{\min} \\
\text{AIC} = -2 \ln(\mathcal{L}) + 2K \\
\label{eq:aic-w}
\end{aligned}
\end{equation}
\noindent

where the AIC value for model $i$ is the negative log-likelihood $\ln(\mathcal{L})$ penalized by the number of estimated parameters $K$. 
We consider three case study species from the Gulf of Alaska (Arrowtooth Flounder, Pacific Ocean Perch, and Pacific Spiny Dogfish) 

\section{Results}

\subsection{Simulation testing}\label{simulation-results}

\begin{figure}[htb]
    \centering
    \includegraphics[width=0.8\linewidth]{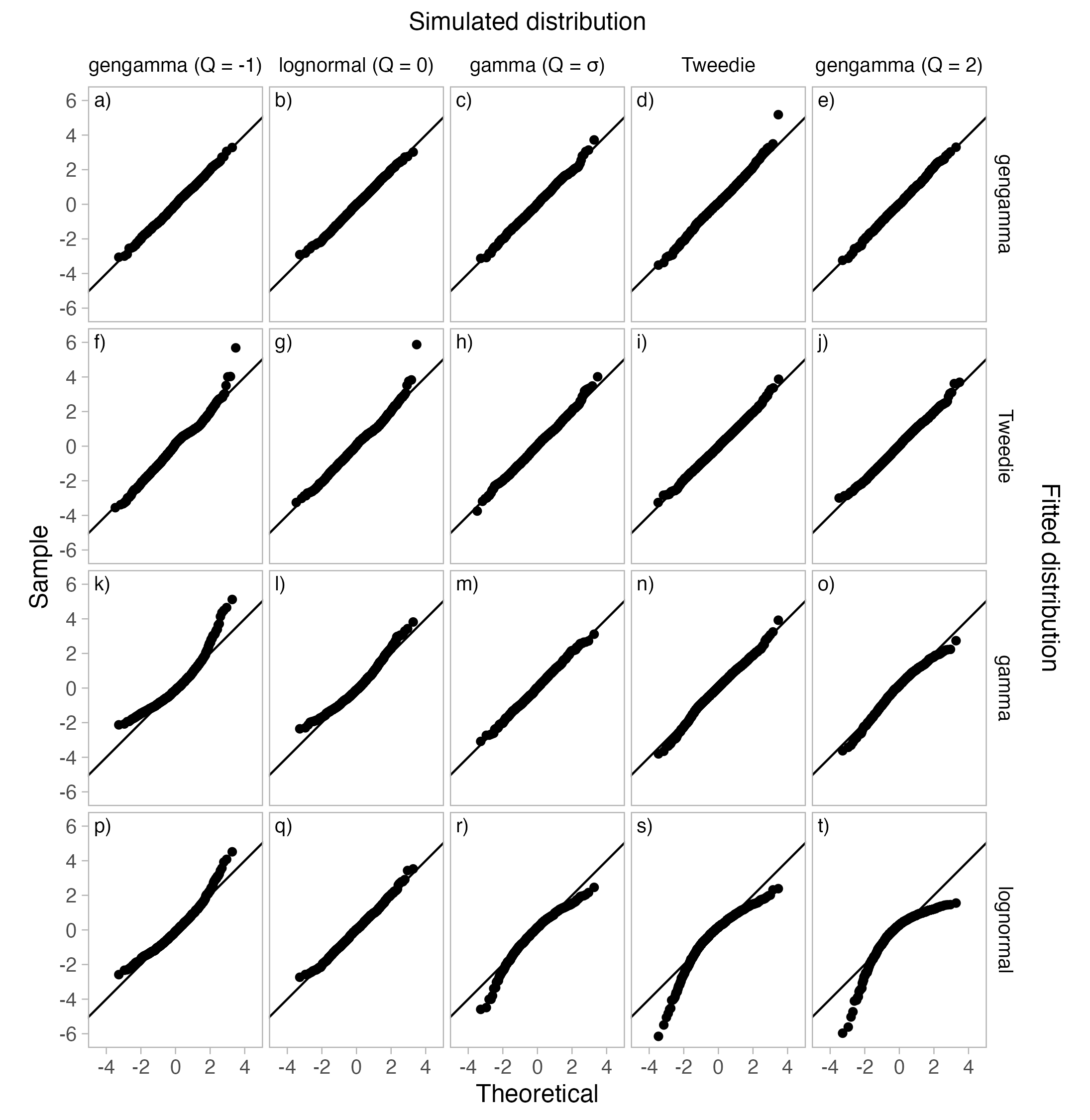}
    \caption{Quantile-quantile (QQ) plot of randomized quantile residuals for example simulated data. Columns correspond to the simulated observation likelihood and rows correspond to the fitted model observation likelihood. Columns are ordered from most heavy-tailed (Q = -1) to most light-tailed (Q = 2). The residuals have been transformed to be normally distributed if the model was consistent with the data. The diagonal line corresponds to the 1:1 line.}
    \label{fig:cross-sim-qq}
\end{figure}

The GGD (``gengamma'' in figures following the R family function name) performed well across the simulated datasets. 
Randomized quantile residuals (RQR) from the models fitted with the GGD had residuals consistent with the expected distribution across the simulations from different distributions (Figure~\ref{fig:cross-sim-qq}). 
The Tweedie distribution also performed well across most of the simulated datasets in terms of the RQR, though it did not perform as well as the GGD when $Q < 0$ (i.e., data that were heavier tailed than the lognormal; Figure~\ref{fig:cross-sim-rqr-supp}). 
Similarly, the gamma distribution performed poorly when $Q < 0$, failing to generate residuals consistent with the model expectations at both the upper and lower tails.
As expected, residuals from models using the lognormal family (Figure~\ref{fig:cross-sim-qq}) indicated over prediction of large values as the tails of the simulated data became lighter. 
For very heavy tails ($Q < -0.5$) the GGD had better residual diagnostics than the lognormal.

\begin{figure}[htb]
    \centering
    \includegraphics[width=1\linewidth]{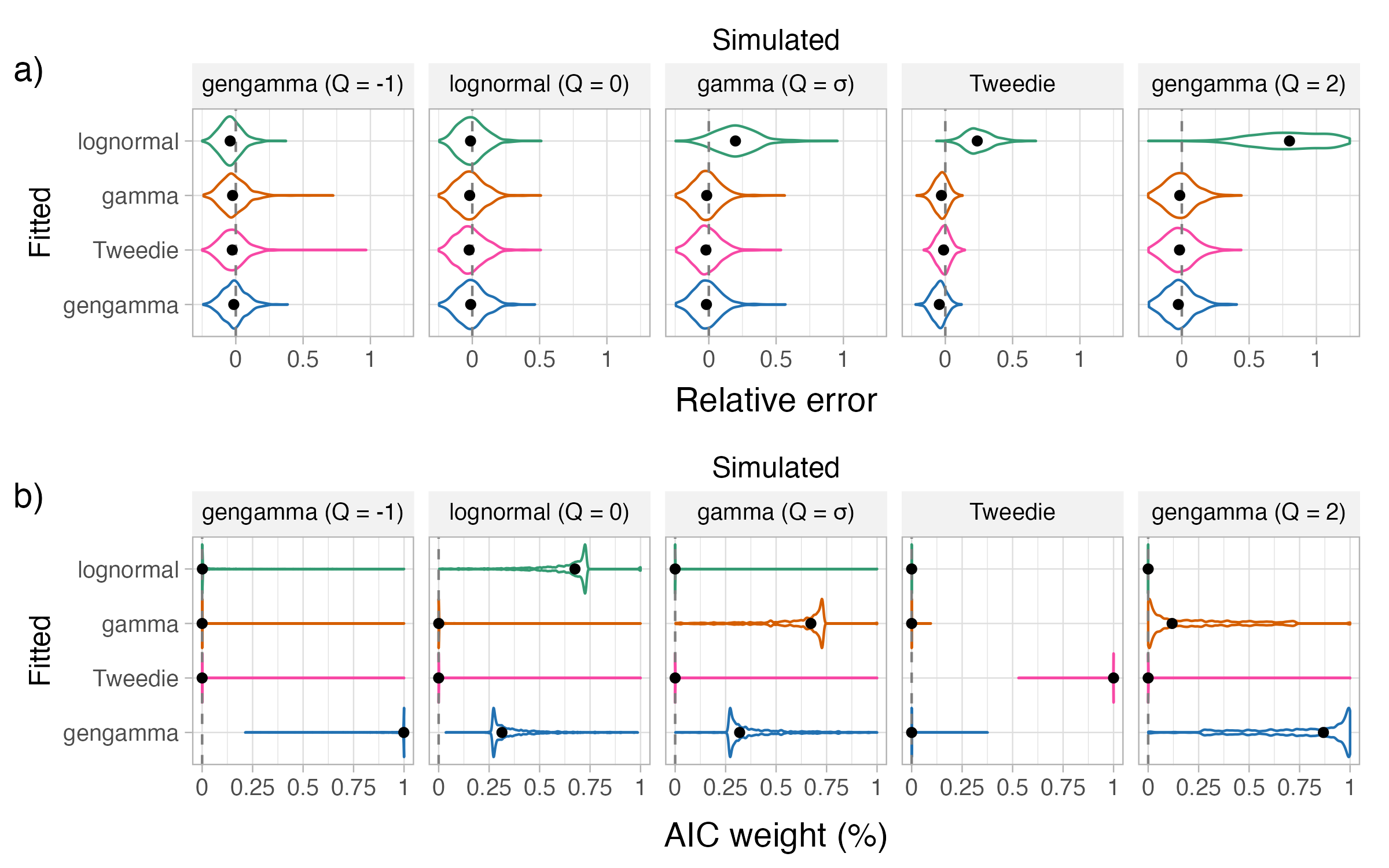}
    \caption{(a) Relative error and (b) AIC weights from 1000 simulation-estimation replicates of biomass density. Black points indicate the median value and violins indicate density. Columns represent the simulated observation likelihood and y-axis labels represent the fitted observation likelihood. Columns are ordered from most heavy-tailed (Q = -1) to most light-tailed (Q = 2). Note that in (a), to improve visual clarity of the median bias near zero, the x-axis has been limited to maximum value of 1, which truncates the right tail of the density for the fitted lognormal relative error.}
    \label{fig:cross-sim-re-aicw}
\end{figure}

The GGD produced unbiased estimates of biomass density and had high accuracy for predicting biomass density across different observation error types. 
The GGD, gamma, and Tweedie families had median relative errors near zero, indicating that indices were not biased regardless of the simulation families used (Figure~\ref{fig:cross-sim-re-aicw}a).
Only the lognormal family had a large relative error when fit to data simulated from $Q > 0$ or the Tweedie distribution (Figure~\ref{fig:cross-sim-re-aicw}a). 
As the simulated data became lighter-tailed (i.e., as $Q$ increased), the mean estimated by the lognormal increased relative to the true mean.

The predictive accuracy of the GGD was robust to the underlying observation process (Figure~\ref{fig:cross-sim-re-aicw}b). 
Across values of $Q$ the GGD had the highest AIC weights, except for the special cases where $Q \to 0$ and $Q = \sigma$ (Figure~\ref{fig:cross-sim-aicw-supp}).
The lognormal ($Q \to 0$) and gamma ($Q = \sigma$) families had the highest AIC weights when fit to their matching simulated data, however the GGD model had median AIC weights of $\sim 31$\% (Figure~\ref{fig:cross-sim-re-aicw}), which is close to the value $\sim 26.8 \% = \frac{e^{-1}}{1 + e^{-1}}$ that is expected when the generalized gamma results in the same marginal likelihood as the corresponding model but has a $\Delta$ AIC that is larger by two units due to the additional parameter $Q$. 

\subsection{Multi-stock analysis}\label{multi-stock-results}

\begin{figure}[htb]
    \centering
    \includegraphics[width=0.9\linewidth]{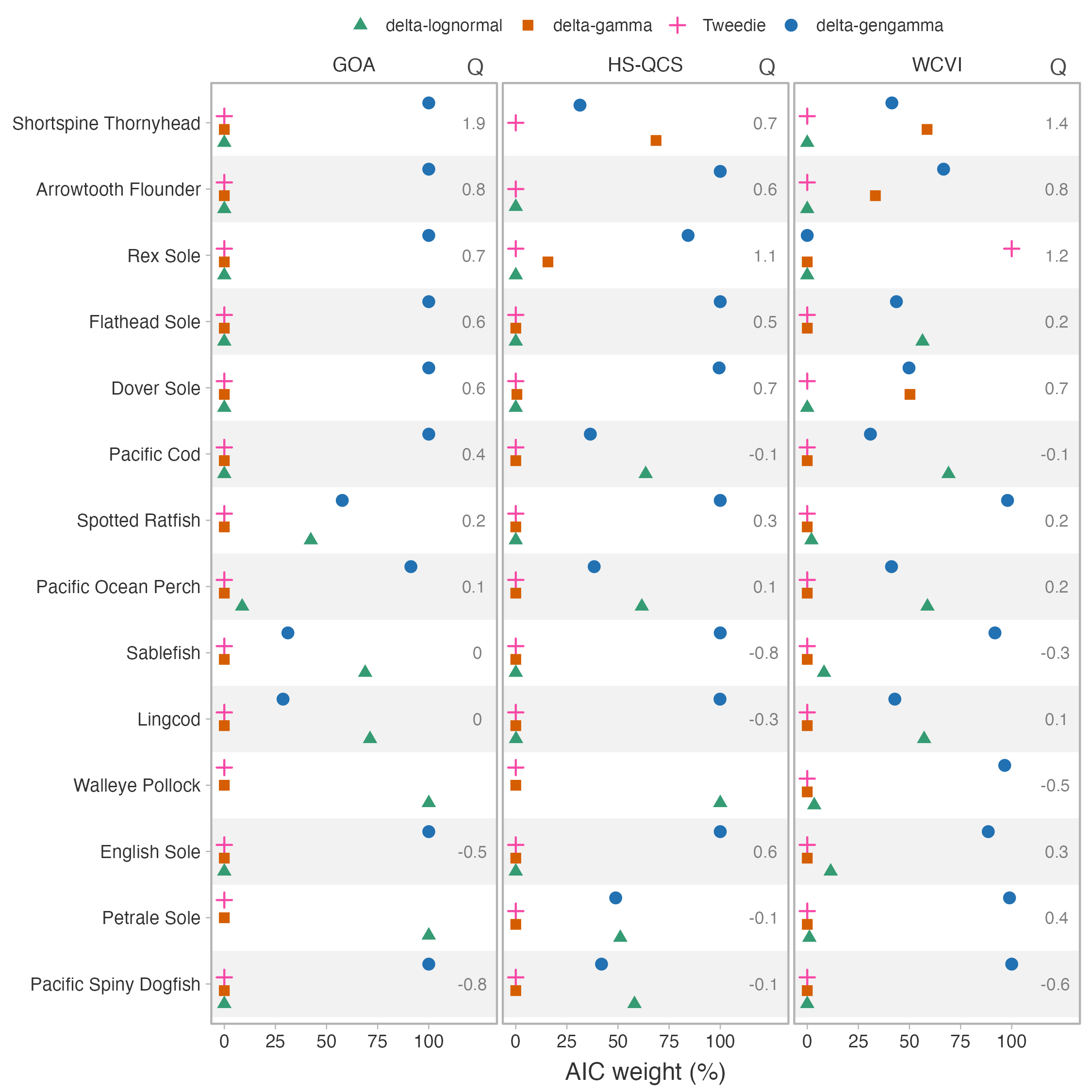}
    \caption{Multi-stock AIC weights. Species are ordered from highest to lowest estimated generalized gamma Q values (shown in text towards the right of each column) within the GOA survey (i.e., from least to most heavy tailed). Where no Q value is indicated, the delta-gengamma model did not converge. GOA: Gulf of Alaska; HS-QCS: Hecate Strait and Queen Charlotte Sound; WCVI: West Coast Vancouver Island.}
    \label{fig:multi-stock-aic}
\end{figure}

Across most species in each region, the delta-GGD was selected as the best or close to best model (Figure~\ref{fig:multi-stock-aic}).
When $Q \to 0$ the AIC weight was split between the lognormal and GGD models, which matches the results of the simulation testing.
At higher values of $Q$, in some species (e.g., Shortspine Thornyhead, Arrowtooth Flounder) the AIC weight was split between the gamma and GGD. 
This happened where the difference between $\sigma$ and the estimated Q was smallest, and in the HS-QCS and WCVI regions, which were less data-rich than the GOA (Figure \ref{fig:sampling-diff-supp}). 

\begin{figure}[htb]
    \centering
    \includegraphics[width=1\linewidth]{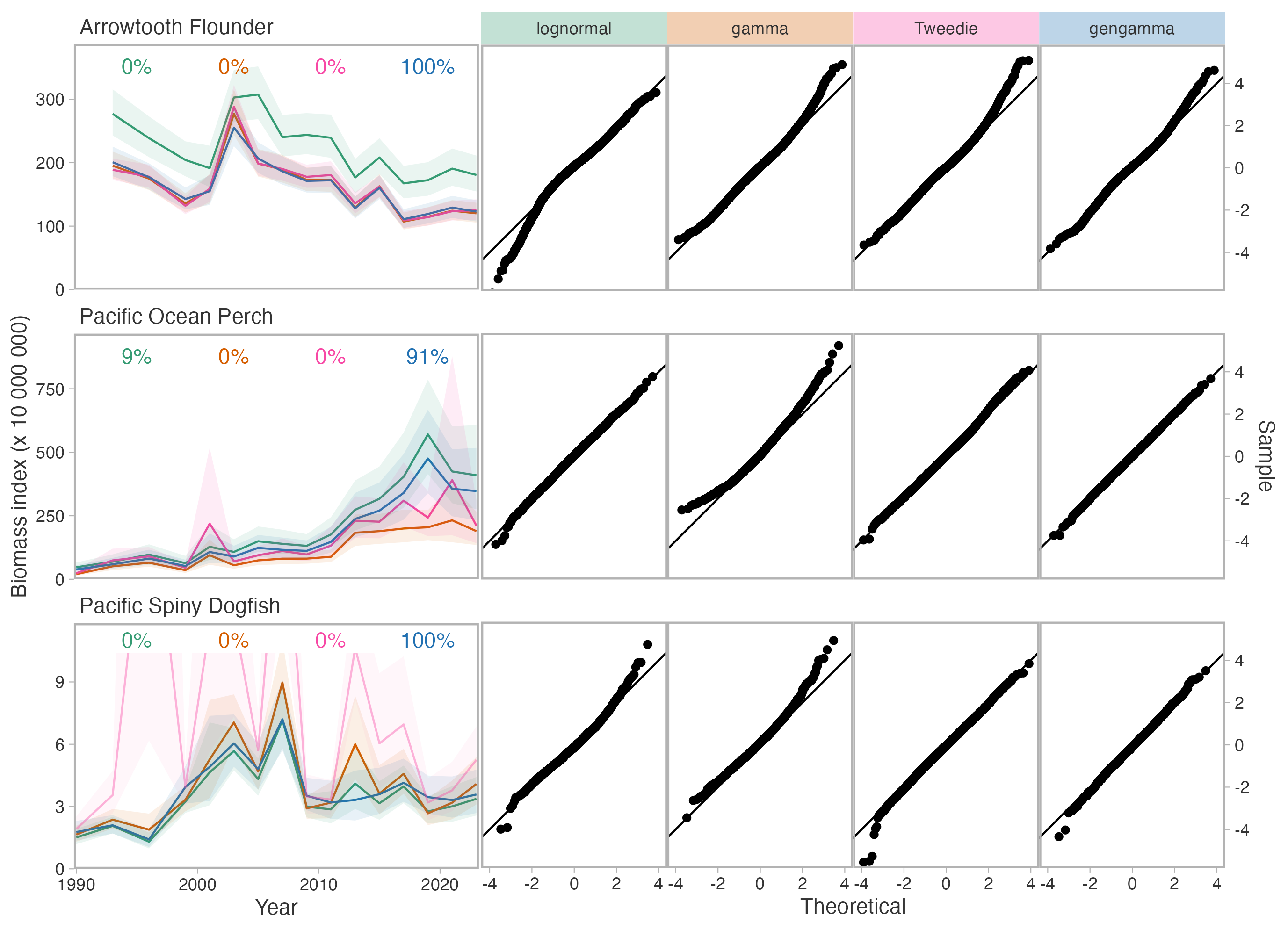}
    \caption{Examples of spatiotemporal-model-derived area-weighted indices of biomass for three groundfish species in the Gulf of Alaska across four observation likelihoods. Left column shows the indices of biomass (mean and 95\% confidence interval) with AIC weight percentages shown at the top. Right columns illustrate associated quantile-quantile plots for randomized quantile residuals that would be distributed as standard normal if the model were consistent with the data. The diagonal line corresponds to the 1:1 line. The panels are ordered from top to bottom in order of decreasing Q value (i.e., light-tailed to heavy-tailed)}
    \label{fig:multi-stock-examples}
\end{figure}

The three examples of spatiotemporal-model-derived area-weighted indices of biomass from the GOA (Arrowtooth Flounder, Pacific Ocean Perch, and Pacific Spiny Dogfish), show that the GGD performed as expected, i.e., resembling the gamma or lognormal distributions depending upon the estimated value for $Q$.
When the estimated $Q$ was high, the index estimated by the delta-GGD model tended to better match the delta-gamma model (e.g., Arrowtooth Flounder in Figure~\ref{fig:multi-stock-examples}). 
As $Q$ approached 0, the index estimated by the GGD model became similar in trend and magnitude as the index estimated by the delta-lognormal model (Figures~\ref{fig:multi-stock-examples}, \ref{fig:index-goa-supp}, \ref{fig:index-hs-qcs-supp}, \ref{fig:index-wcvi-supp}).
The scale estimated by the GGD was similar to the design-based scale for Arrowtooth Flounder (Table \ref{model-design-ratio}). 
However, for Pacific Ocean Perch and Pacific Spiny Dogfish, although the GGD had the highest AIC weight, the scale of the delta-gamma was closer to the design-based index (Table \ref{model-design-ratio}).

\clearpage

\section{Discussion}

We present a parameterization of the GGD in terms of the mean and coefficient of variation, which allows its incorporation into existing software for spatiotemporal modelling. 
In a simulation experiment, the GGD demonstrated robust performance against distribution misspecification and when tested across 15 species in three regions the GGD had the highest or close to the highest AIC weight in most cases. 
Compared to other commonly used distributions (e.g., lognormal, gamma, Tweedie), the GGD offers a flexible alternative for fitting fisheries-independent survey observations.

Across a range of simulated $Q$ values and when fit to data simulated from other distributions, the GGD performed as expected in the simulation analysis.
The GGD closely resembled the lognormal distribution for lognormal-simulated data and the gamma distribution for gamma-simulated data. 
In contrast, bias increased as simulated $Q$ increased (tails became lighter than lognormal) when the lognormal distribution was fitted.
The GGD offers greater flexibility by fitting the lognormal distribution when appropriate but avoiding the bias seen with the lognormal model as tail behaviour diverges from lognormal.
Moreover, the GGD's ability to estimate distributions that are heavier tailed than the lognormal ($Q < 0$), with tails between the lognormal and gamma ($\sigma > Q > 0 $), or with tails lighter than the gamma ($Q > \sigma$) adds a new and practical option for use in comparative studies \citep[e.g.,][]{dick2004, thorson2021scale} and real-world index-standardization.
Lack-of-fit in the tails of the distribution when incorrectly fitting the gamma and lognormal distributions was clearly seen in quantile-quantile plots. These plots therefore appear useful to diagnose instances when the GGD might be appropriate.

In practice, the GGD demonstrated high predictive accuracy across the species and regions we studied. The GGD performed similarly to, or better than, the lognormal and gamma models.
This suggests that the extra parameter does not lead to overfitting or overparameterization in most cases.
Furthermore, the GGD indicates that for some species such as Pacific Spiny Dogfish, a distribution that is more heavy-tailed than the lognormal is more appropriate for trend estimation.
The risk of selecting the lognormal or gamma distribution for assessing fish populations may be heightened by unpredictable demographic changes over time, and selecting one of these distributions at a point in time may prevent the detection of such changes in the future. 
From our results, it seems prudent for fisheries researchers to use the GGD instead which could estimate a different $Q$ as new data are added. In some cases, it may be useful to allow $Q$ to vary through time such as by fitting separate GGD parameters for each year of the survey to detect whether there are shifts in the shape of tails over time.

Despite the additional flexibility of the GGD, in cases where index scale is important, like when the catchability coefficient is fixed a priori in an assessment-model, it is important to compare the model-based index scale with a design-based scale or to a priori use the gamma or Tweedie distributions if a design-based index is unavailable. 
In our analysis, we show that for GOA Pacific Ocean Perch and GOA Pacific Spiny Dogfish AIC and the estimated GGD $Q$ parameter identify the lognormal or a distribution that is more heavy-tailed than lognormal, respectively, as parsimonious.
However, for both stocks the index estimated using a gamma distribution more closely matches the design-based index. 
Therefore, we re-iterate the caution from \citet{thorson2021scale} that specific purposes might be better suited by a priori model selection and avoid using AIC as the only criterion for model performance.

We note that using the mean-CV parameterization of the GGD has many potential applications outside of the spatiotemporal model presented here. 
For example, \citet{albertsen2017} compared the GGD against other likelihoods when fitting abundance-at-age or alternative age-composition data. 
However, that study used the \citet{prentice1974} parameterization (i.e., using a central tendency that does not match the distribution mean). 
Existing R packages, such as flexsurv, parameterize the GGD using the location ($\mu$), scale ($\sigma$) and shape ($Q$) parameters, whereas the packages VAST and sdmTMB use the mean-CV parameterization shown in Table \ref{tab:table1}.
We recommend that further applications consider using the mean-CV parameterization, so that model parameters (i.e., estimated local density for index-standardization models, or estimated abundance-at-age for stock-assessment models) are estimated to match the mean of data (rather than a central tendency that is difficult to interpret and depends on details of the parameterization). 
For example, \citet{cadigan2001} explored the gamma and lognormal for sequential population models, and \citet{jiao2004} explored these options for modelling variation in recruitment. 
Both of these applications could potentially benefit from testing the GGD. 
Additionally, \citet{cacciapaglia2024} explored the generalized-gamma distribution for index standardization, using the implementation available in VAST that we document for the first time in the present paper. 
That study found the GGD was selected by AIC for three of four stocks, and had $\Delta \mathrm{AIC} < 2$ for the remaining stock compared with the selected lognormal distribution. 
Finally, Monnahan et al. (In review) has used the GGD to weight indices in stock assessment models, again using the parameterization derived from the present study.

We recommend future research test the performance of SDMs using the GGD for other applications such as characterizing ecosystem drivers and projecting spatial distributions under alternative climate scenarios.
We also recommend further comparative research to explore biological attributes that are associated with survey catches that have larger outliers than predicted by the lognormal (i.e., $Q < 0$). 
Previous studies have called these ``extreme catch events'' (ECEs) and have represented ECEs using a mixture distribution that arises from a mixture of dispersed and shoaling habitats \citep{thorson2012AgentBased} or with multivariate-t random fields \citep{anderson2019swans}.
These ECEs appear to arise predictably for aggregating and shoaling species (e.g., some rockfishes, dogfish), and indices for these species then typically have substantially higher uncertainty.
We therefore recommend that the GGD be used as a generalized method for identifying biological characteristics associated with these ECEs.

\bibliography{refs}

\section{Acknowledgments}

We thank C.C. Monnahan, who identified the potential role of the generalized gamma distribution for index standardization while contributing to the Discussion section of \citet{thorson2021scale}. We thank K. Kristensen and the developers of Template Model Builder, without which this work would not be computationally feasible. We thank C.N. Rooper and L.A.K. Barnett for helpful comments that improved this manuscript.

\clearpage

\renewcommand{\thefigure}{S\arabic{figure}}
\renewcommand{\thetable}{S\arabic{table}}
\setcounter{figure}{0}
\setcounter{table}{0}
\setcounter{section}{0}
\setcounter{subsection}{0}
\setcounter{subsubsection}{0}
\setcounter{page}{1}
\setcounter{equation}{0}
\setcounter{secnumdepth}{0} 
\linenumbers
\resetlinenumber

\clearpage

\begin{Center}
\section{Supporting Information}
\end{Center}

\begin{figure}[htb]
    \centering
    \includegraphics[width=0.8\linewidth]{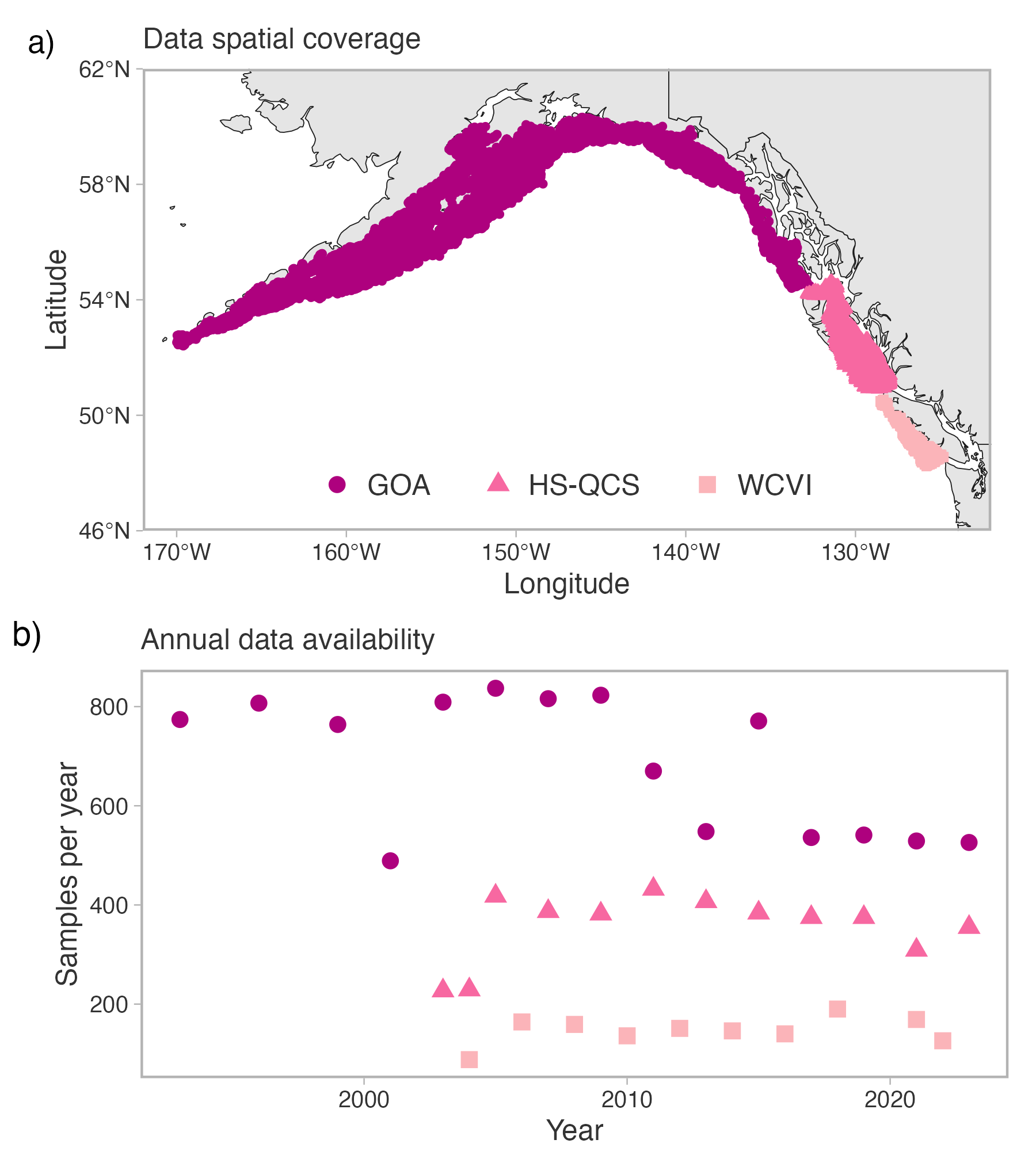}
    \caption{(a) Spatial coverage and (b) annual data availability of the three survey regions used in the multi-stock analysis: Gulf of Alaska (GOA), Hecate Strait and Queen Charlotte Sound (HS-QCS), and West Coast Vancouver Island (WCVI).}
    \label{fig:sampling-diff-supp}
\end{figure}

\begin{landscape}

\begin{figure}[htb]
    \centering
    \includegraphics[width=\linewidth]{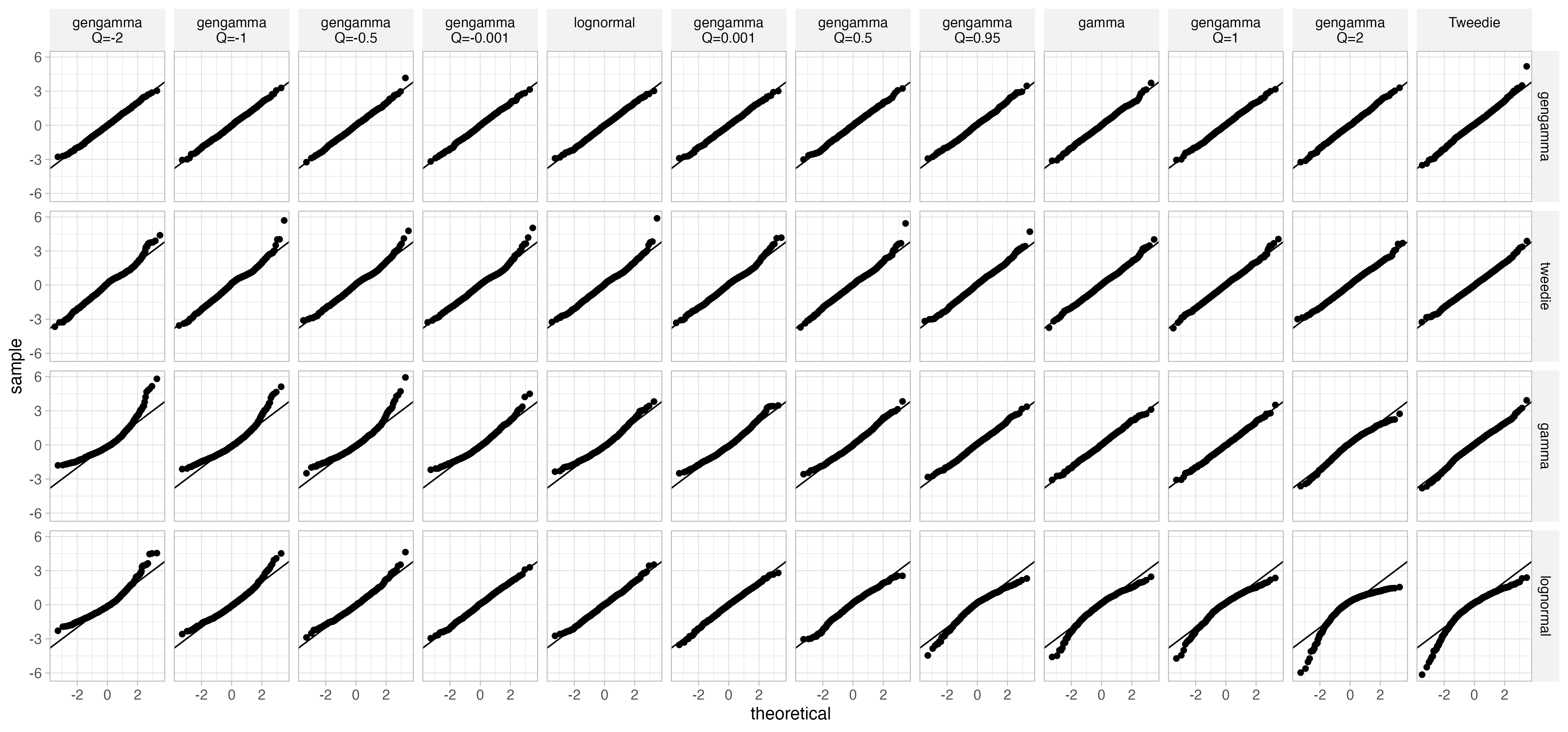}
    \caption{Quantile-quantile (QQ) plot of randomized quantile residuals for example simulated data. Columns correspond to the simulated observation likelihood and rows correspond to the fitted model observation likelihood. Columns are ordered from most heavy-tailed (Q = -1) to most light-tailed (Q = 2). The residuals have been transformed to be normally distributed if the model was consistent with the data. The diagonal line corresponds to the 1:1 line. This is an extended version of Figure~\ref{fig:cross-sim-qq}.}
    \label{fig:cross-sim-rqr-supp}
\end{figure}

\begin{figure}[htb]
    \centering
    \includegraphics[width=1\linewidth]{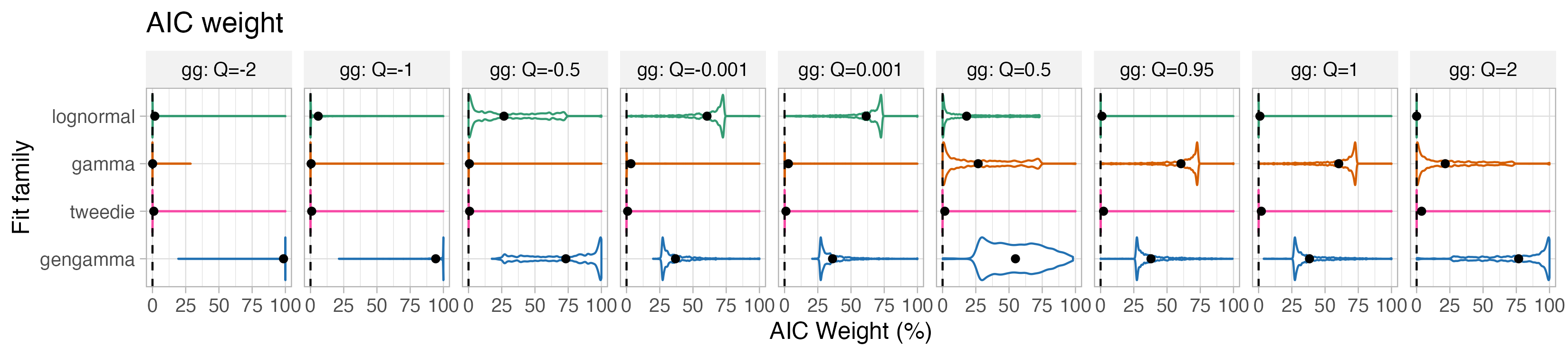}
    \caption{AIC weights from 1000 simulation-estimation replicates of biomass density. Black points indicate the median value and violins indicate density. Columns represent the simulated observation likelihood from the generalised gamma distribution (gg) and y-axis labels represent the fitted observation likelihood. Columns are ordered from most heavy-tailed (Q = -1) to most light-tailed (Q = 2). This is an extended version of Figure~\ref{fig:cross-sim-re-aicw}b.}
    \label{fig:cross-sim-aicw-supp}
\end{figure}

\end{landscape}

\begin{figure}[htb]
    \centering
    \includegraphics[width=1\linewidth]{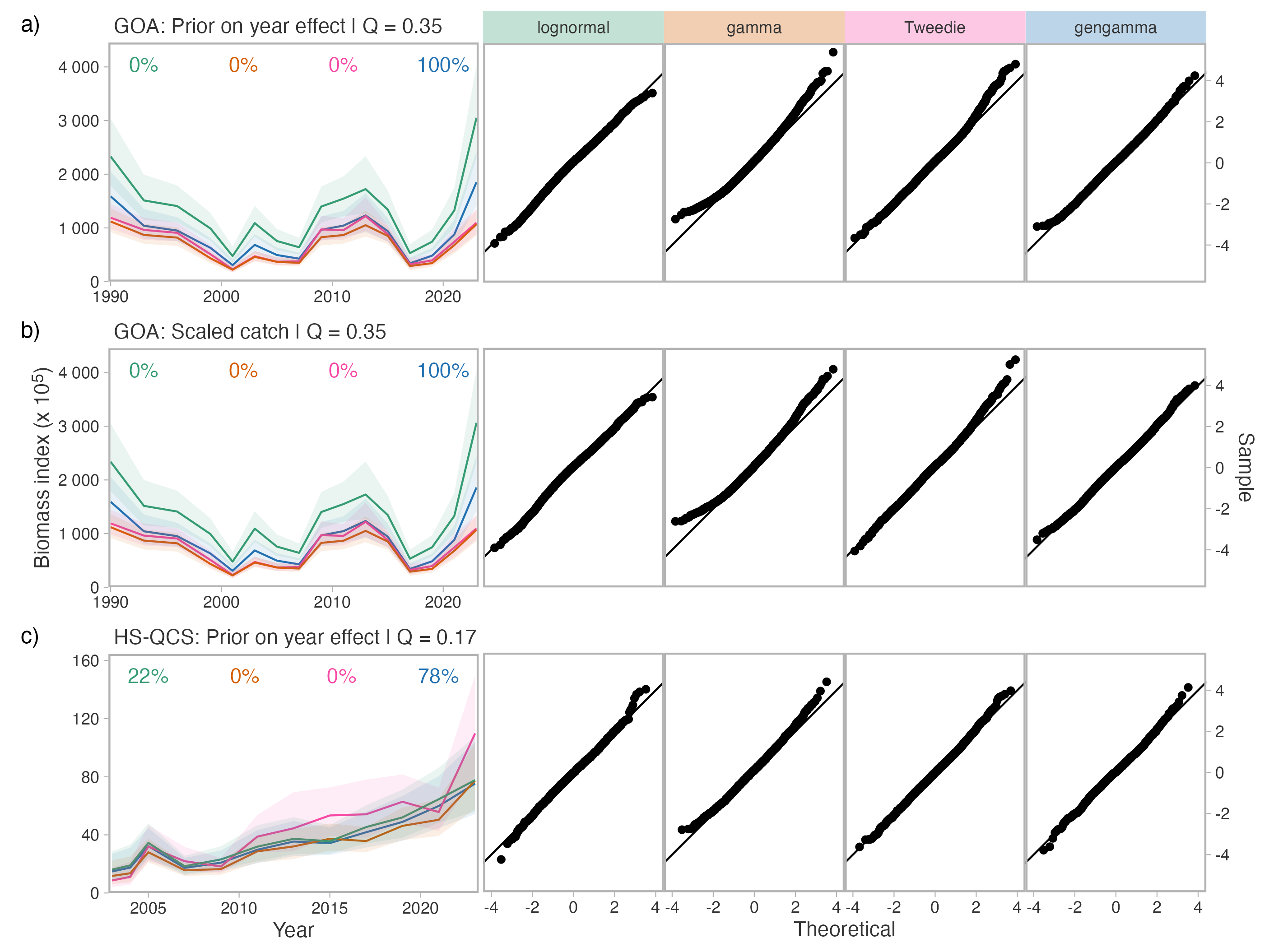}
    \caption{Spatiotemporal-model-derived area-weighted indices of biomass for Walleye Pollock in the Gulf of Alaska (GOA) and Hecate Strait \& Queen Charlotte Sound (HS-QCS) across four observation likelihoods. 
    Left column shows the biomass indices (mean and 95\% confidence interval) estimated either (a, c) using a weak prior that constrained the year estimates, where $\alpha_{t} \sim \operatorname{N}(0, 30^{2})$, or (b) by scaling the catch variable by 1 / 100 kg. 
    Right columns illustrate associated quantile-quantile plots for randomized quantile residuals that would be distributed as standard normal if the model were consistent with the data. The diagonal line corresponds to the 1:1 line.}
    \label{fig:pollock-index-rqr-supp}
\end{figure}

\begin{figure}[htb]
    \centering
    \includegraphics[width=1\linewidth]{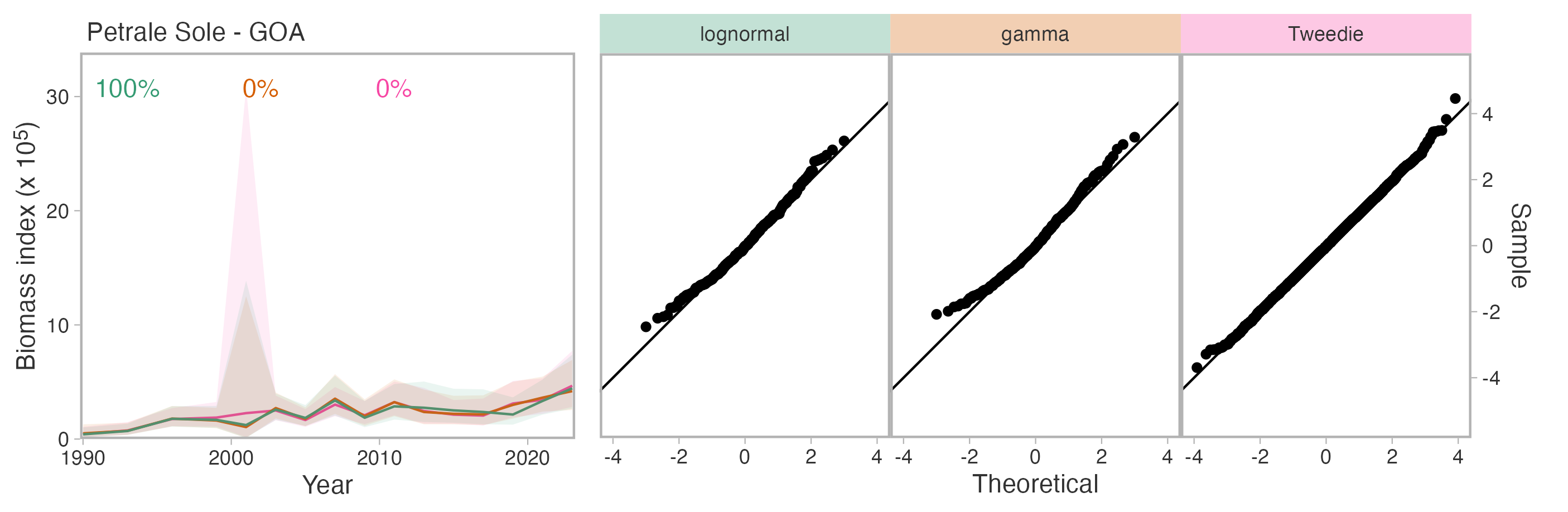}
    \caption{Spatiotemporal-model-derived area-weighted indices of biomass for Petrale Sole in the Gulf of Alaska (GOA).
    Left column shows the biomass indices (mean and 95\% confidence interval).
    Right columns illustrate associated quantile-quantile plots for randomized quantile residuals that would be distributed as standard normal if the model were consistent with the data. The diagonal line corresponds to the 1:1 line.
    The GGD model is not shown because it did not converge.}
    \label{fig:petrale-index-rqr-supp}
\end{figure}

\begin{figure}[htb]
    \centering
    \includegraphics[width=1\linewidth]{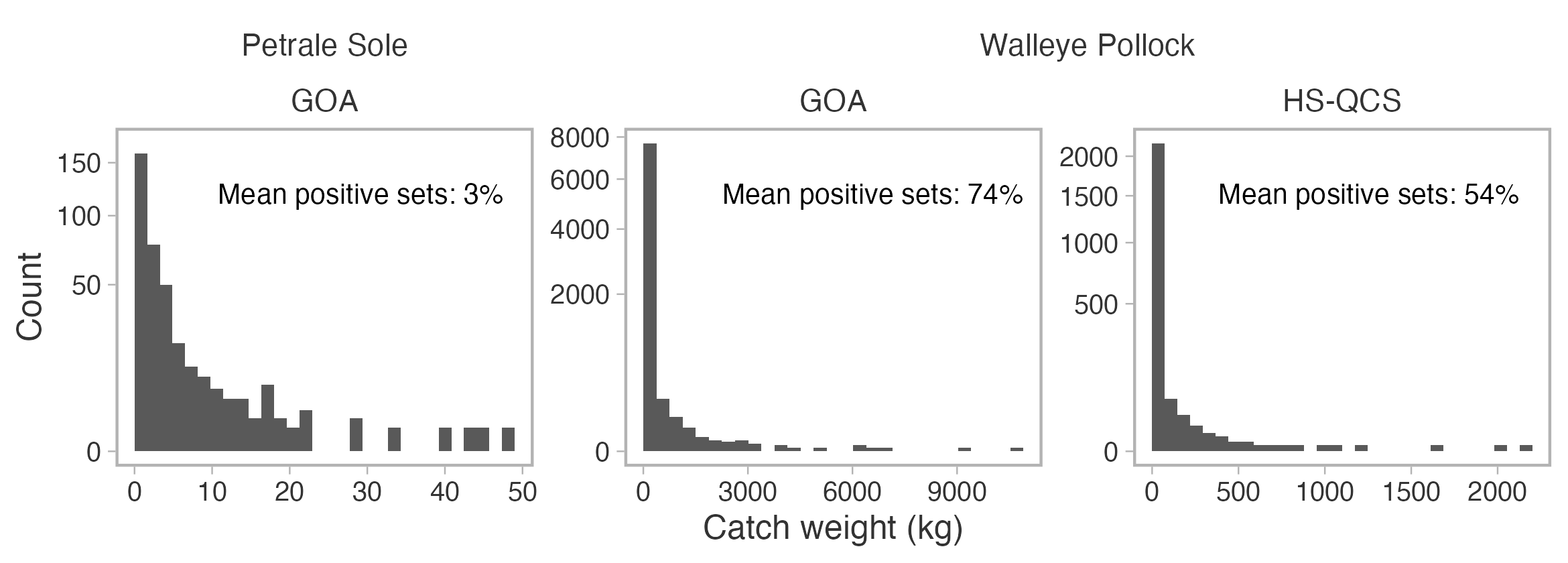}
    \caption{Frequency of positive catches weights (kg) observed for three species-region combinations (Petrale Sole-GOA, Walleye Pollock-GOA, Walleye Pollock-HS-QCS). The mean proportion of survey sets where these species were encountered (catch weight $>$ 0) is shown on each panel. These three species-regions did not converge when the delta-gengamma was fit in the main analysis. Note the y-axes are square-root transformed to improve visualisation of the tails of the data.}
    \label{fig:trouble-species-supp}
\end{figure}

\begin{figure}[htb]
    \centering
    \includegraphics[width=1\linewidth]{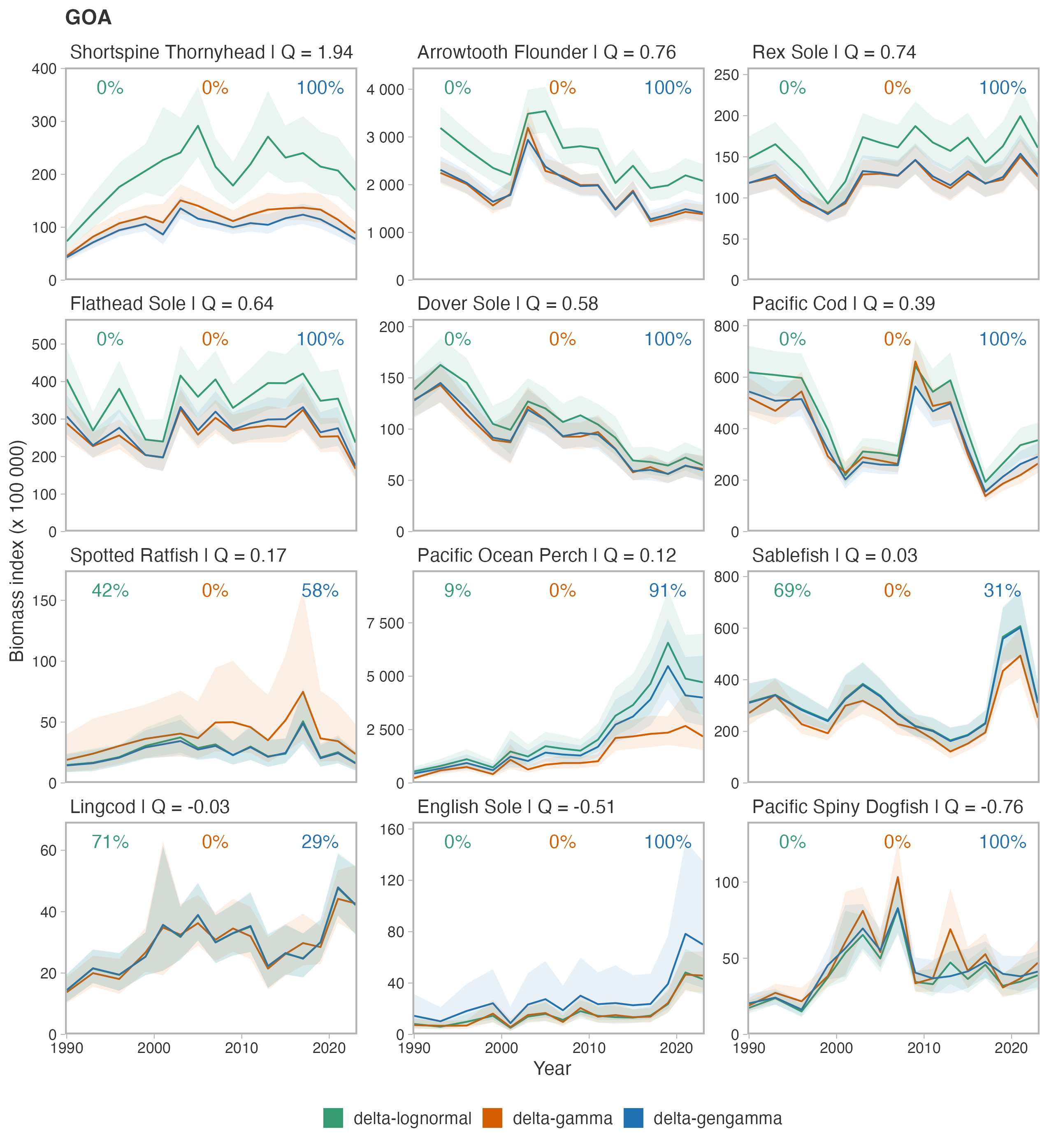}
    \caption{Spatiotemporal-model-derived area-weighted indices of biomass (mean and 95\% confidence interval) for twelve groundfish species in the Gulf of Alaska (GOA) across three observation likelihoods.
    Only species for which the delta-gengamma model converged are shown. 
    Indices estimated using the Tweedie family were omitted to improve visualisation of the differences in the indices estimated using the delta-lognormal (green), delta-gamma (orange), and delta-gengamma (blue). 
    The AIC weight is indicated for each family at the top of each panel; these values match those in Figure \ref{fig:multi-stock-aic}. 
    The panels are ordered from top to bottom, left to right, in order of decreasing Q value (i.e., light-tailed to heavy-tailed).}
    \label{fig:index-goa-supp}
\end{figure}

\begin{figure}[htb]
    \centering
    \includegraphics[width=1\linewidth]{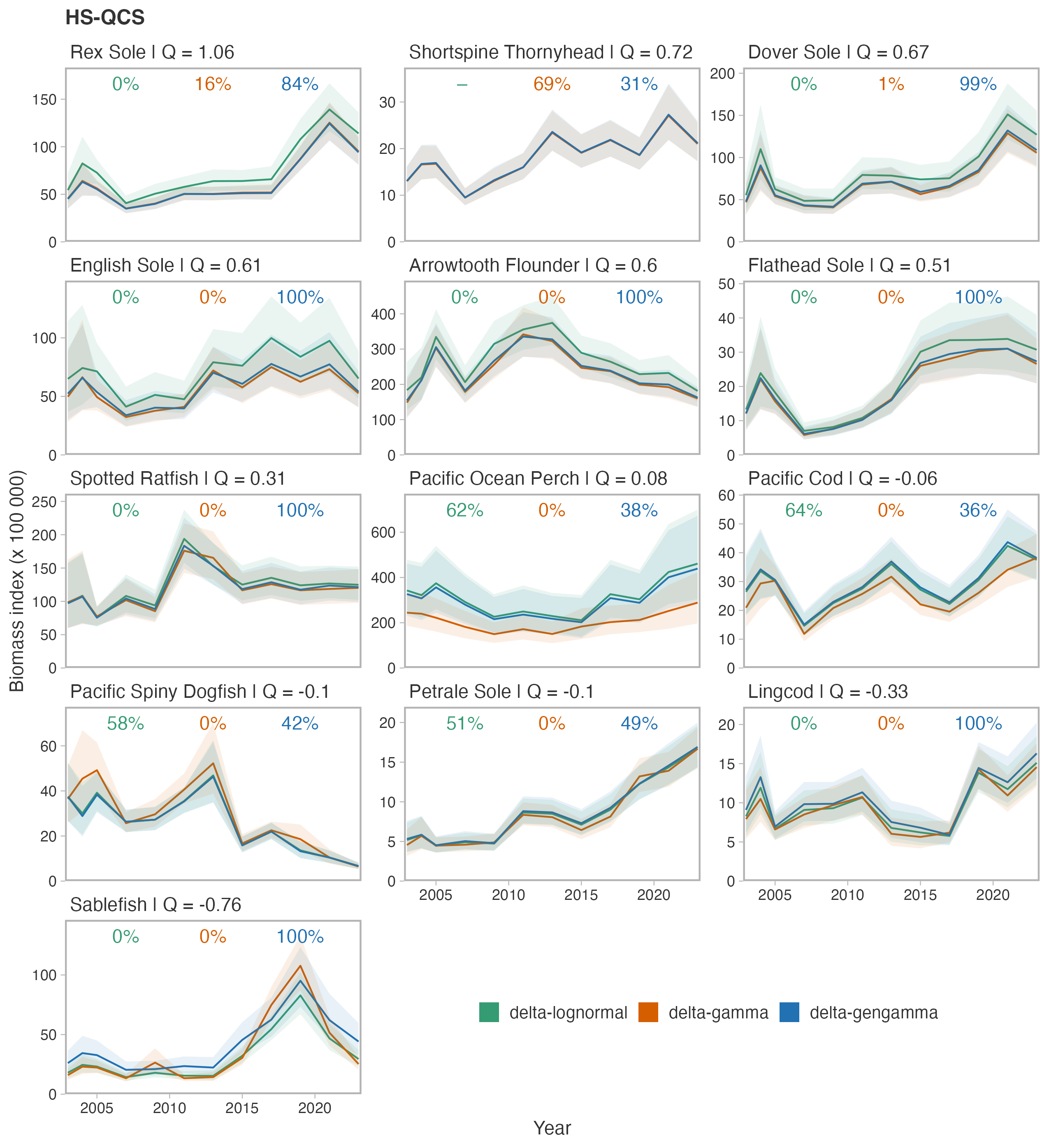}
    \caption{Spatiotemporal-model-derived area-weighted indices of biomass (mean and 95\% confidence interval) for thirteen groundfish species in the Hecate Strait and Queen Charlotte Sound (HS-QCS) across three observation likelihoods. See description of Supplemental Figure \ref{fig:index-goa-supp} for details. There is no AIC weight for Shortspine Thornyhead because this model did not converge, see Figure \ref{fig:multi-stock-aic}.}
    \label{fig:index-hs-qcs-supp}
\end{figure}

\begin{figure}[htb]
    \centering
    \includegraphics[width=1\linewidth]{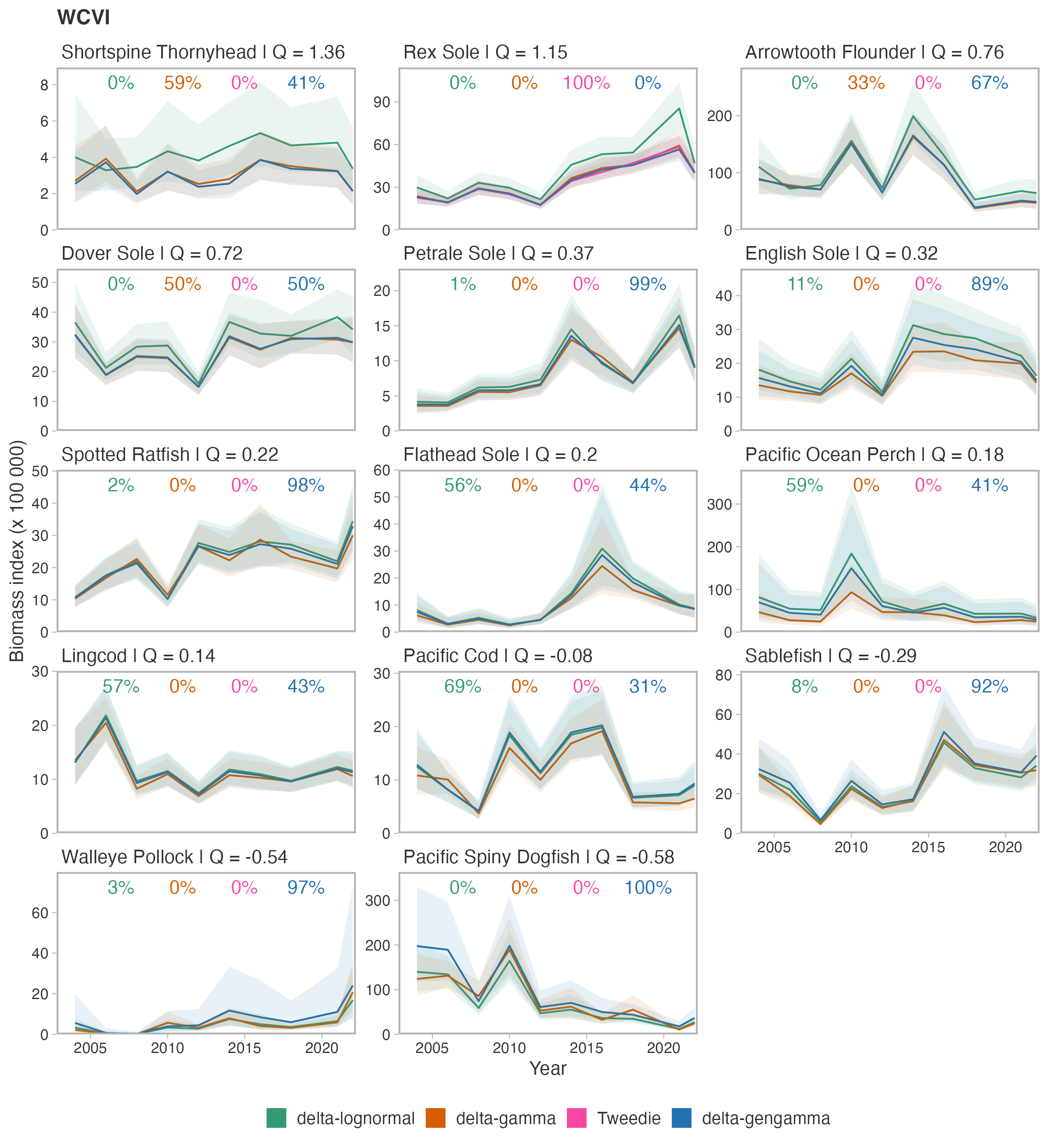}
    \caption{Spatiotemporal-model-derived area-weighted indices of biomass (mean and 95\% confidence interval) for fourteen groundfish species in the Hecate Strait and Queen Charlotte Sound (HS-QCS) across three observation likelihoods. See description of Supplemental Figure \ref{fig:index-goa-supp} for details. One difference from Figure \ref{fig:index-goa-supp} is that the biomass index estimated using the Tweedie is shown on the Rex Sole panel because it was the best fitting model (Figure \ref{fig:multi-stock-aic}).}
    \label{fig:index-wcvi-supp}
\end{figure}

\begin{table}[h!]
\centering
\caption{Ratios of mean estimated biomass from modelled indices relative to design-based indices, ordered by decreasing difference from 1 (i.e., a value of 1 indicates the scale of the model-based index equals the scale of the design-based index).}
\label{model-design-ratio}
\begin{tabular}{llc}
\toprule
Species & Family & Ratio \\
\midrule
\multirow{4}{*}{Arrowtooth Flounder} 
  & delta-gamma     & 1.04 \\
  & delta-gengamma  & 1.04 \\
  & Tweedie         & 1.05 \\
  & delta-lognormal & 1.43 \\
\cmidrule(lr){1-3}
\multirow{4}{*}{Pacific Ocean Perch} 
  & delta-gamma     & 1.28 \\
  & Tweedie         & 1.78 \\
  & delta-gengamma  & 2.04 \\
  & delta-lognormal & 2.43 \\
\cmidrule(lr){1-3}
\multirow{4}{*}{Pacific Spiny Dogfish} 
  & delta-gamma     & 0.72 \\
  & Tweedie         & 1.33 \\
  & delta-gengamma  & 0.66 \\
  & delta-lognormal & 0.62 \\
\bottomrule
\end{tabular}
\end{table}

\clearpage{}

\section{GGD Model Convergence}\label{ggd-convergence}

For three species-region combinations (Walleye Pollock-GOA, Walleye Pollock-HS-QCS, Petrale Sole-GOA), the delta-GGD model did not converge (Figure~\ref{fig:multi-stock-aic}).
In the case of Walleye Pollock, we improved estimation for both the GOA and HS-QCS regions by including a weak prior that constrained the year estimates, where $\alpha_{t} \sim \operatorname{N}(0, 30^{2})$. 
Scaling the catch variable by 0.01 kg allowed convergence of the GOA stock (Figure~\ref{fig:pollock-index-rqr-supp}), which suggests that estimation of the GGD may be sensitive to scale and/or the starting values used for $\sigma$. 
However, scaling by the catch weight (kg) by 1 / 10, or 1 / 100 did not enable convergence of the delta-GGD model for the HS-QCS stock. 
In the refit models, the delta-GGD model had the highest AIC weight in both the GOA and HS-QCS stocks (Figure~\ref{fig:pollock-index-rqr-supp}). 
Surprisingly, the $\hat{Q}$ estimate for the GOA Walleye Pollock was 0.35 in the GOA even though in the main analysis, where only the delta-lognormal, delta-gamma, and Tweedie were compared, the delta-lognormal was selected as the best model (AIC weight = 100\%, Figure \ref{fig:multi-stock-aic}).
Furthermore, the scale of the estimated index was more similar to that estimated by the delta-gamma and Tweedie. 
Neither applying weak priors nor scaling the response enabled convergence for Petrale Sole in the GOA, likely because the Petrale Sole had only a 3\% encounter rate in the GOA (Figure~\ref{fig:trouble-species-supp}).
Although the other three models for GOA Petrale Sole converged, the index estimate for 2000 had a high uncertainty and high mean CV on the index (CV = 0.39, Figure~\ref{fig:petrale-index-rqr-supp}).

\end{document}